\documentclass[12pt,preprint]{aastex}

\newcommand{\be}{\begin{equation}}
\newcommand{\ee}{\end{equation}}
\newcommand{\pindex}{{ p }} 
\newcommand{\ibar}{{ \langle I \rangle }} 

\usepackage{graphicx}
\usepackage{epstopdf}
\usepackage{natbib}

\slugcomment{{\it Astrophysical Journal Letters, in Press}}
\shorttitle{Anomalous Planetary Radii}
\shortauthors{Laughlin, Adams \& Crismani}

\begin{document}

\title{On the Anomalous Radii of the Transiting Extrasolar Planets }

\author{Gregory Laughlin \altaffilmark{1} and Matteo Crismani\altaffilmark{1}}
\altaffiltext{1}{UCO/Lick Observatory, 
Department of Astronomy and Astrophysics, 
University of California at Santa Cruz,
Santa Cruz, CA 95064}

\author{Fred C. Adams \altaffilmark{2}}
\altaffiltext{2}{Department of Physics, University of Michigan,
    Ann Arbor, MI 48109}

\newcommand{\superscript}[1]{\ensuremath{^{\textrm{#1}}}}
\newcommand{\subscript}[1]{\ensuremath{_{\textrm{#1}}}}

\begin{abstract}
We present a systematic evaluation of the agreement between the observed radii of 90 well-characterized transiting extrasolar giant planets and their corresponding model radii. Our model radii are drawn from previously published calculations of core-less giant planets that have attained their asymptotic radii, and which have been tabulated for a range of planet masses and equilibrium temperatures. (We report a two-dimensional polynomial fitting function that accurately represents the models). As expected, the model radii provide a statistically significant improvement over a null hypothesis that the sizes of giant planets are completely independent of mass and effective temperature. As is well known, however, fiducial models provide an insufficient explanation; the planetary {\it radius anomalies}, ${\cal R} \equiv R_{\rm obs}-R_{\rm pred}$, are strongly correlated with planetary equilibrium temperature. We find that the radius anomalies have a best-fit dependence, ${\cal R}\propto T_{\rm eff}^{\alpha}$, with $\alpha=1.4\pm0.6$. Incorporating this relation into the model radii leads to substantially less scatter in the radius correlation. The extra temperature dependence represents an important constraint on theoretical models for Hot Jupiters. Using simple scaling arguments, we find support for the hypothesis of Batygin and Stevenson (2010) that this correlation can be attributed to a planetary heating mechanism that is mediated by magnetohydrodynamic  coupling between the planetary magnetic field and near-surface flow that is accompanied by ohmic dissipation at adiabatic depth. Additionally, we find that the temperature dependence is likely too strong to admit kinetic heating as the primary source of anomalous energy generation within the majority of the observed transiting planets.
\end{abstract}

\keywords{Stars: Planetary Systems, Planets and Satellites: General}

\section{Introduction}\label{sec:intro}

The disparate radii of the transiting extrasolar planets have defied a straightforward explanation. With nearly 100 well-characterized examples now known, it is clear that where Jovian planets are concerned, variations in mass and stellar insolation are responsible only for a fraction of the observed range in planetary sizes.

Radius anomalies emerged with the discovery of the first transiting extrasolar planet, HD 209458b \citep{Charbonneau00}. Structural evolutionary models computed by, e.g., \cite{Bodenheimer01} and \cite{Guillot02} suggested that a H-He dominated planet with HD 209458b's mass, insolation, and age should have a radius of order $R\sim1.1R_{\rm Jup}$. This figure is startlingly at odds with the observed value, $R=1.38\pm0.02R_{\rm Jup}$ \citep{Southworth10}.

A decade of additional discoveries has made it clear that HD 209458b's radius anomaly is by no means anomalous, and the status of discussion in the field is well-covered in the recent reviews by \cite{Showman10}, \cite{BurrowsOrton10}, and \cite{Baraffe10}. The extraordinary variation in observed radii in the planetary population is indicated by Figure 1. This diagram charts the measured radii and uncertainties for the 90 transiting planets with accurately measured masses against the planets' orbit-averaged effective temperatures, given by 
\begin{equation}
T_{\rm eff}=\left({R_{\star}\over{2a}}\right)^{1/2} {T_{\star} \over{(1-e^{2})^{1/8}}} \, ,
\end{equation}
where $R_{\star}$ is the stellar radius, $a$ is the semi-major axis, $e$ is the orbital eccentricity, $T_{\star}$ is the stellar effective temperature, and zero albedos have been assumed. Only planets with $0.1\,M_{\rm Jup}<M_{\rm pl}<10.0\,M_{\rm Jup}$ were used.\footnote{All observational data taken from www.exoplanet.eu, accessed 12/03/2010.}

It is evident that the observationally well-characterized transiting planets exhibit a wide range of radii for a given mass, and a number of models have been advanced to explain this dispersion. Planets that are smaller than expected, with HD 149026b \citep{Sato05} providing the canonical example, have generally had their small radii attributed to a large fraction of heavy elements in their interiors. For planets such as HD 209458b, which are larger (and sometimes much larger) than expected, explanatory models must generally appeal to a cryptic source of heating at adiabatic depth. A number of heating mechanisms have been proposed, including dissipation stemming from tidal orbital circularization \citep{Bodenheimer01}, ``kinetic heating" in which wind energy is converted into heat \citep{Guillot02}, and Ohmic dissipation \citep{Batygin10}. This letter shows that current observational evidence indicates that the radius anomaly, the difference between the observed radius and that predicted by theoretical models, displays a well-defined correlation with temperature (from equation [1]). Further, this temperature dependence can be interpreted to favor a magnetohydrodynamic mechanism in which Ohmic heating and attendant magnetic braking of the velocity field at sufficient atmospheric ionization fraction play significant roles. Our approach relies on scaling arguments, as we believe that the extant observations are not yet sufficient to permit the discrimination between detailed models containing multiple free parameters.

\section{A Comparison Between Structural Models and the Observations}\label{sec:intro}

Naively, one might expect that the radius of a mature gas-giant planet is primarily determined by its mass and by the amount of radiative energy that it receives from its star. This conjecture can be tested by evaluating model radii, $R_{m_{i}}$, of ``baseline'' evolutionary models for H-He composition planets spanning a range of masses and insolation, and comparing with those of known corresponding transiting planets, $R_{o_{i}}$. If the model has explanatory power for the aggregate of $N$ known planets, then it should produce a statistically significant decrease in the quantity,
\begin{equation}
\chi^{2}_{m}={1\over{(N-N_{f})}}{\displaystyle\sum\limits_{i=1}^{N} {({R_{m_{i}}-R_{o_{i}})^{2}}\over{\sigma_{i}^{2}}} }\,\,\,\, ,
\end{equation}
in comparison to $\chi^{2}_{null}$, obtained by replacing $R_{m_{i}}$ with $R_{av}=1.2R_{\rm Jup}$, the average observed radius for transiting planets having $0.1 M_{\rm Jup}<M_{\rm pl}<10.0 M_{\rm Jup}$. In the above equation, $N_f=2$ is the number of free parameters ($M_{\rm pl}$, $T_{\rm eff}$) in the explanatory model.

Our baseline models were published by \cite{Bodenheimer03}; hereafter BLL. As described in BLL, radius estimates, $R_{m_{i}}$, were computed with a Henyey-type planetary structure calculation, and the reader is referred to that paper for details regarding the input physics and assumptions. The model radii were tabulated at 4.5 Gyr for a grid of assumed $M_{\rm pl}$ and $T_{\rm eff}$. We ignore the small dependence of radius on age for mature planets and use bilinear interpolation to obtain an estimate for $R_{m_{i}}$ at given $M_{\rm pl}$ and $T_{\rm eff}$. Our estimates are drawn from BLL's model sequence of core-free solar-composition planets with no anomalous energy sources.

For the 90 transiting planets, we find $\chi^{2}_{m}=23.5$ and  $\chi^{2}_{null}=32.6$. Not surprisingly, this result indicates that the baseline structural models can explain some, but by no means all, of the the observed variation in planetary radii. As an alternative to bilinear interpolation between table values, it can be useful to have a simple fitting relation, $R_{\rm pl}(M_{\rm pl}, T_{\rm eff})$. Defining $m=\log_{10}(M_{\rm pl}/M_{\rm Jup})$, and $t=T_{\rm eff}/1000$, we find that the two-dimensional polynomial fitting function,
\begin{eqnarray}
R_{\rm pl}/R_{\rm Jup} =1.08417+0.0940857\,m-0.242831\,m^{2}+0.0947349\,m^{3} && \nonumber \\
 +0.0387851\,t+0.00243981\,mt-0.0244656\,m^{2}t +0.0130659\,m^{3}t && \nonumber \\
 +0.0240409\,t^{2}-0.0419296\,mt^{2}+0.00693348\,m^{2}t^{2}+0.00302157\,m^{3}t^{2} \, ,
\end{eqnarray}
provides an acceptable approximation to the BLL baseline structural models throughout the region where $0.1 M_{\rm Jup}<M_{\rm pl}<10.0 M_{\rm Jup}$, and $100<T_{\rm eff}<2500$.\footnote{Applying this polynomial relation in place of bilinear interpolation yields $\chi^{2}_{m'}=18.30$ for the aggregate of 90 transiting planets.}

For each planet, we define a radius anomaly, ${\cal R}_{i}=R_{m_{i}}-R_{o_{i}}$, and look for correlations between $\cal R$ and other measured quantities (such as $T_{\rm eff}$, $T_{\star}$, $M_{\star}$, etc). Many authors, (e.g., Enoch et al. 2010) have noticed that planetary radii tend to swell dramatically with increasing insolation. Figure 2 illustrates the significant correlation between $\cal R$ and planetary $T_{\rm eff}$. Using a bootstrap replacement method \citep{Press92}, we find a best fit power-law dependence, 
\begin{equation}
{\cal R}\propto T_{\rm eff}^{1.4\pm0.6}\, .
\end{equation}

Among the various mechanism that have been invoked to explain the radius anomalies, we expect that both Ohmic heating \citep{Batygin10} and kinetic heating \citep{Guillot02} should show a positive correlation between $\cal R$ and $T_{\rm eff}$. We can ask, furthermore, whether the measured exponent, $\alpha=1.4\pm0.6$, is consistent with either of these proposed mechanisms. 

In the treatment of \cite{Batygin10}, energy deposition in the planet is approximated by integrating over the resistivity in each mass element
\begin{equation}
{\dot{E}}= \int \int \int {{\bf J}^{2}\over{\sigma_r(r)}} dV\, ,
\end{equation}
where $\bf J$ is the current density and $\sigma_r(r)$ is the conductivity in the near-surface (but still adiabatic) layers where the resistive heating occurs. For order-of-magnitude purposes, they ignore the radial dependence, $\sigma_r(r) \sim \sigma_r$. Further, it is assumed that the relevant currents are induced by the planet's intrinsic dipole field, ${\bf B}$, so that
\begin{equation}
{\bf J}=\sigma_r({\bf v}\times {\bf B}) \,  \quad {\rm and}  \quad \dot{E}\propto \sigma_r |{\bf v}|^{2}|{\bf B}|^{2} \, .
\end{equation}
We want to estimate the temperature dependence, $\dot{E}(T_{\rm eff})$, implied by the Batygin-Stevenson model. First, we need an estimate of how wind kinetic energy at adiabatic depth scales with temperature. One possibility is that there is a constant partitioning between thermal energy and kinetic energy, an assumption that conforms with the kinetic heating hypothesis of \cite{Guillot02}, which posits that a fraction, $\eta$, of the total flux received by a planet is converted into kinetic energy by atmospheric pressure gradients, and that this energy is dissipated at depth by Kelvin-Helmholtz instabilities. In this case, one has $v\propto T_{\rm eff}^{2}$. Alternately, one can examine three-dimensional hydrodynamical models for irradiated giant planets that experience a range of insolations. These models suggest a much weaker dependence of wind speed on $T_{\rm eff}$ at large optical depths where energy dissipation can effectively couple to the planetary
 structure. For example, the \cite{Lewis10} model for Gliese 436b ($M_{\rm pl}=24 M_{\oplus}$, $T_{\rm eff}=707\,K$) which adopts a solar metallicity atmosphere, finds $v\sim100 \, {\rm m s}^{-1}$ at $P=10\,{\rm bar}$ depth. This wind velocity is nearly identical to the average wind velocity at 10 bar depth found by \cite{Showman09} for a solar metallicity atmospheric model of HD 189733 ($M=1.2 M_{\rm Jup}$, $T_{\rm eff}=1202\,K$).
For our rough analysis, we choose $ v\propto T_{\rm eff}^{1/2}$, bearing in mind that the temperature's power-law index might reasonably fall anywhere in the range from 0 to 2. 

Second, we follow \cite{Batygin10} and assume that the {\it Elsasser Number}, $\Lambda = {\sigma_i {\bf B}^{2}/{2 \rho \Omega}}$, is roughly constant across the aggregate of transiting planets.
(The quantity $\sigma_i$ is the conductivity appropriate to the interior regions where the ${\bf B}$ field is assumed to be generated by dynamo action, $\rho$ is the density of the planet, and $\Omega$ is its angular spin frequency). To the precision of our analysis, this relation holds for the bounding cases of Jupiter and the Sun; Jupiter has a magnetic field roughly ten times the solar value, and has a spin period that is roughly 100 times shorter. Both bodies have similar densities and are highly conductive in their interiors. With $\Lambda$ and $\sigma_i$ assumed constant, we can remove the magnetic field dependence, yielding $\dot{E}\propto \sigma_r \rho \Omega T_{\rm eff} \,$. If we further assume synchronous rotation (which should hold for hot Jupiters on circular orbits), then ${1/{\Omega}} \propto a^{3/2}$ and $a \propto T_{\rm eff}^{-2}$ (via equation [1]), giving $\dot{E}\propto \sigma_r T_{\rm eff}^{4}$.

We need to identify how the dissipation-layer conductivity, $\sigma_r$, depends on temperature. The conductivity is directly tied to the ionization fraction, which in turn is determined by the Saha equation
\begin{equation}
{n^{+}_j n_e \over n_j - n^{+}_j}  = 
\left( {m_e kT \over 2 \pi \hbar^2} \right)^{3/2} {2 g^{+}_{j}\over{g_{j}}}
\exp \left[ - I_j /kT \right] \, , 
\end{equation}
where the $j$ chemical species typically include Na, K, Li, Rb, Fe, Cs,
and Ca, and $T$ is the local temperature.  The quantity $I_j$ is the ionization potential (for the first
electron to be ionized) and ranges from 4.3 eV
(for K) to 7.9 eV (for Fe), whereas $g^{+}_{j}$, and $g_{j}$ are the degeneracies of states for the ions and neutrals, respectively, of species $j$. Note that $n_e$ is the total electron
number density (since all electrons can interact with each species of
ion).  In the relevant temperature regime, ionization fractions are low, so that $n_j^{+} \ll n_j$. Summing over all species and simplifying yields
\begin{equation}
n_e = \left( {m_e kT \over 2 \pi \hbar^2} \right)^{3/4} 
\left[ \sum_j f_j n_j \exp \left[ - I_j / kT \right] \right]^{1/2} 
\, , 
\end{equation}
where $f_j = {2 g_j^{+} / g_j}$.

Next, we note that the electrical conductivity, $\sigma_r$, appropriate to the dissipative region has the form 
\begin{equation}
\sigma_r = {n_e \over n} {e^2 \over m_e A} 
\left( {\pi m_e \over 8 k T} \right)^{1/2} \, , 
\end{equation}
where $A$ is a weighted cross section, see, e.g. \cite{Tipler02}. The quantity of interest 
is the index $\pindex$ defined by 
\begin{equation}
\pindex = {T \over \sigma_r} {\partial \sigma_r \over \partial T} = 
- {1 \over 2} + {T \over n_e} {\partial n_e \over \partial T} \, .  
\end{equation}
The index $p$ allows us to estimate the temperature dependence of the conductivity, $\sigma_r$. 
Using our expression for $n_e$, we obtain 
\begin{equation}
\pindex = {1 \over 4} + {T \over 2} 
\left[ \sum_j f_j n_j \exp \left[ - I_j / kT \right] \right]^{-1} 
{\partial \over \partial T} 
\left[ \sum_j f_j n_j \exp \left[ - I_j / kT \right] \right] \, .  
\end{equation}
After differentiation, this equation can be written in the (apparently) simpler form  
\begin{equation}
\pindex = {1 \over 4} + { \ibar \over 2 kT} \, , 
\end{equation}
where we have defined 
\begin{equation}
\ibar \equiv 
\left[ \sum_j f_j n_j \exp \left[ - I_j / kT \right] \right]^{-1} 
\left[ \sum_j I_j f_j n_j \exp \left[ - I_j / kT \right] \right] \, .  
\end{equation}
Since $kT \ll I_j$, the exponential suppression factors dominate the
behavior of this function. As a result, we can consider the largest 
term to dominate (which will be the term with the smallest ionization
potential $I_\ast$): 
\begin{equation}
\pindex \approx {1 \over 4} + {I_\ast \over 2 kT} 
\end{equation}
For $T$ = $T_{eff}$ = 2000 K, the quantity $2kT$ = 0.345 eV. If $I_\ast$ = 4 eV, 
then $\pindex \approx 12$. It's clear that $\dot{E}(T_{\rm eff})$ depends sensitively on $T_{\rm eff}$, and that for a given choice of $I_\ast$, the index $\pindex$ increases sharply with $T_{\rm eff}$ (for $I_\ast$ = 4 eV, $\pindex$=15.7 at 1500 K, and $\pindex$=9.5 at 2500 K).

A similar result is obtained by 
Batygin \& Stevenson (2010), who present a simple but accurate model   
for the conductivity as a function of radius, where 
$\sigma \propto \exp [ - (r-R)/H]$, where $H$ is the conductivity 
scale height (which is about twice the thermal scale height). 
Since $H \sim T$, the index $\pindex \approx (r-R)/H$, 
which is of order 10.  

Next, we need to establish the dependence of the radius anomaly, $\cal R$ on the internally generated power, $\dot{E}$.  This depends on the planetary mass and the size of the assumed core -- it's easier to inflate a low-mass planet without a core. For a typical hot Jupiter, such as HAT-P-13, models based on the BLL prescription indicate that by increasing $\dot{E}$ more than tenfold, from $3.5\times10^{25} \, {\rm erg\,  s^{-1}}$ to $5.6\times10^{26} \, {\rm erg\, s^{-1}}$, one inflates the radius from $R=1.2 R_{\rm Jup}$ to $R=1.36 R_{\rm Jup}$ \citep{Batygin09}. This single example suggests that $\cal R$ depends on $\dot{E}$ raised to a small fractional power, $x$. \cite{Gu04} have computed sequences of evolutionary models (again using a method similar to that described in BLL) which systematically explore how planetary radii depend on internally generated power. We obtained a fit to their sequences and find that an index $x\sim1/6$ applies over a wide range of masses for planets having $\log(R_{\rm pl}/R_{\rm Jup})<0.3$. This regime covers nearly all of the known transiting cases. Using $\pindex$=12, we arrive at a predicted dependence
\begin{equation}
{\cal R}\propto T^{\alpha} \, {\approx}\, T_{\rm eff}^{{\,(12+4)/{6}}} \approx T_{\rm eff}^{\,2.666}\,\,\,\,,
\end{equation}
for which the index $\alpha$ is more that one power of temperature steeper than suggested by the currently observed aggregate of transiting planets (${\cal R} \propto T^{1.4}$).

Caution regarding the quantitative aspect of the dependence given by equation (15) is in order.  Order-of-magnitude scaling arguments can at best give only rough insight into the physical balance that determines the radii
of the hot Jupiters. As mentioned above, there are considerable uncertainties associated with the wind speeds at adiabatic depth, and the treatment leading to equation (14) is highly idealized. We have assumed, furthermore, that the Ohmic heating rate is directly proportional to the conductivity, $\sigma_r$, in the weather layer. If the ionization fraction is large enough, a more fully self-consistent treatment must account for the back-reaction onto the wind velocity, ${\bf v}$, generated by the Lorentz force

\begin{equation}
\rho {d{\bf v}\over{dt}} = {1\over{c}} \sigma_r({\bf v}\times{\bf B})\times{\bf B}\, .
\end{equation}

Indeed, \cite{Perna10a} and \cite{Perna10b} have suggested that the bulk Lorentz force acts as a drag term on the velocity field. In this event, higher temperatures increase ionization, generating Lorentz forces that brake the velocity field. Reduced wind-speeds lower the rate of Ohmic dissipation, which in turn would decrease the temperature power-law index $\alpha$. A complete understanding of this process requires a full 3D MHD approach. While such simulations are not beyond current computational capabilities, the lack of detailed knowledge of the atmospheric conditions likely renders them somewhat premature at this time.

\section{Discussion}

Transiting extrasolar planets tend to have radii that are larger than those predicted by structural models parameterized by $M_{\rm pl}$ and $T_{\rm eff}$. The resulting radius anomalies, ${\cal R}$, are strongly correlated with effective temperature, with ${\cal R}\propto T_{\rm eff}^{1.4}$ providing a maximum reduction in variance. The simple scaling arguments of the previous section suggest that Ohmic heating may play a significant role, and that further progress may be made as self-consistent MHD-mediated weather models are constructed.

A very rapid increase in $\dot{E}$ with temperature poses difficulties for some posited explanations for the radius anomalies. For example, the \cite{Guillot02} kinetic heating mechanism proposes that a constant fraction, $\eta \sim 0.01$, of the total flux received by a planet is converted into kinetic energy by atmospheric gradients, and that this energy is dissipated at adiabatic depth. While the flow patterns on the surfaces of strongly irradiated extrasolar planets are still a matter of debate, a $\dot{E}\propto T_{\rm eff}^{4}$ scaling is implied by the kinetic heating hypothesis, leading to ${\cal R}\propto T^{2/3}$, which (to $> 1\sigma$ confidence) is weaker than the observed dependence. This argument does not imply that kinetic heating is absent, rather, the implication is that it is unlikely to be the primary mechanism that determines the radius anomalies.

Finally, we note that substantial variance in the observed radii remains  after the  ${\cal R}\propto T_{\rm eff}^{1.4}$ scaling has been removed. Inclusion of the best-fit $\beta T_{\rm eff}^{\alpha}$ dependence into the model radii increases $N_{f}$ from 2 to 4, while causing $\chi^{2}_{m}$ as measured with Equation [2] to decrease from $\chi^{2}_{m}=23.5$ to $\chi^{2}_{m'}=14.8$. Clearly, processes other than Ohmic heating are also contributing. For example, our base-line models unrealistically assume solar-composition planets. Both Jupiter and Saturn are substantially super-solar, and the small radii of planets such as HD 149026b and CoRoT 8b can only be understood if elements heavier than H/He contribute more than 50\% of the planetary mass. Figure 3 indicates that there is evidence for a significant correlation between the residual radius anomalies and host star metallicities. This correlation is generally interpreted as evidence that metal-rich protoplanetary disks lead to planets with larger core masses. Furthermore, a fraction of the known transiting planets have eccentric orbits, meaning that interior heating from tidal orbital circularization must be ongoing in some of the observed planets. For modest eccentricities, and at a given time-average tidal quality factor, $Q$, tidal heating rates scale with $R_{p}^{5}$ \citep{MD99}. This strong dependence could be a determining factor in producing the residual radius anomalies for many of the observed planets. Furthermore, even if a planet's orbital eccentricity is currently zero, it may still be inflated as a consequence of thermal inertia from a tidal heating episode that occurred in the past \citep{Ibgui09}.

The number of transiting planets with well-determined properties will increase rapidly in the coming years. An expanded catalog of planets, in conjunction with improvements to the structural models, will allow empirical quantities, such as the power-law index ``$\alpha$" that we've used here, to be determined with increased confidence. In addition, follow-up campaigns from platforms such as JWST will improve our understanding of the atmospheric conditions on short-period planets, and we can look forward to a time when the construction of detailed, fully self-consistent 3D MHD climate models are a sensible response to the quality and depth of the observational data sets.

\acknowledgments
The authors acknowledge useful discussions with Konstantin Batygin, and we are grateful to an anonymous referee for a prompt and insightful report. This work was funded by NASA grant NNX08AY38A and NASA/Spitzer/JPL grant 1368434 to GL, and by NASA grant NNX07AP17G to FCA.


\newpage

\begin{figure}
{
 \includegraphics[scale= .4]{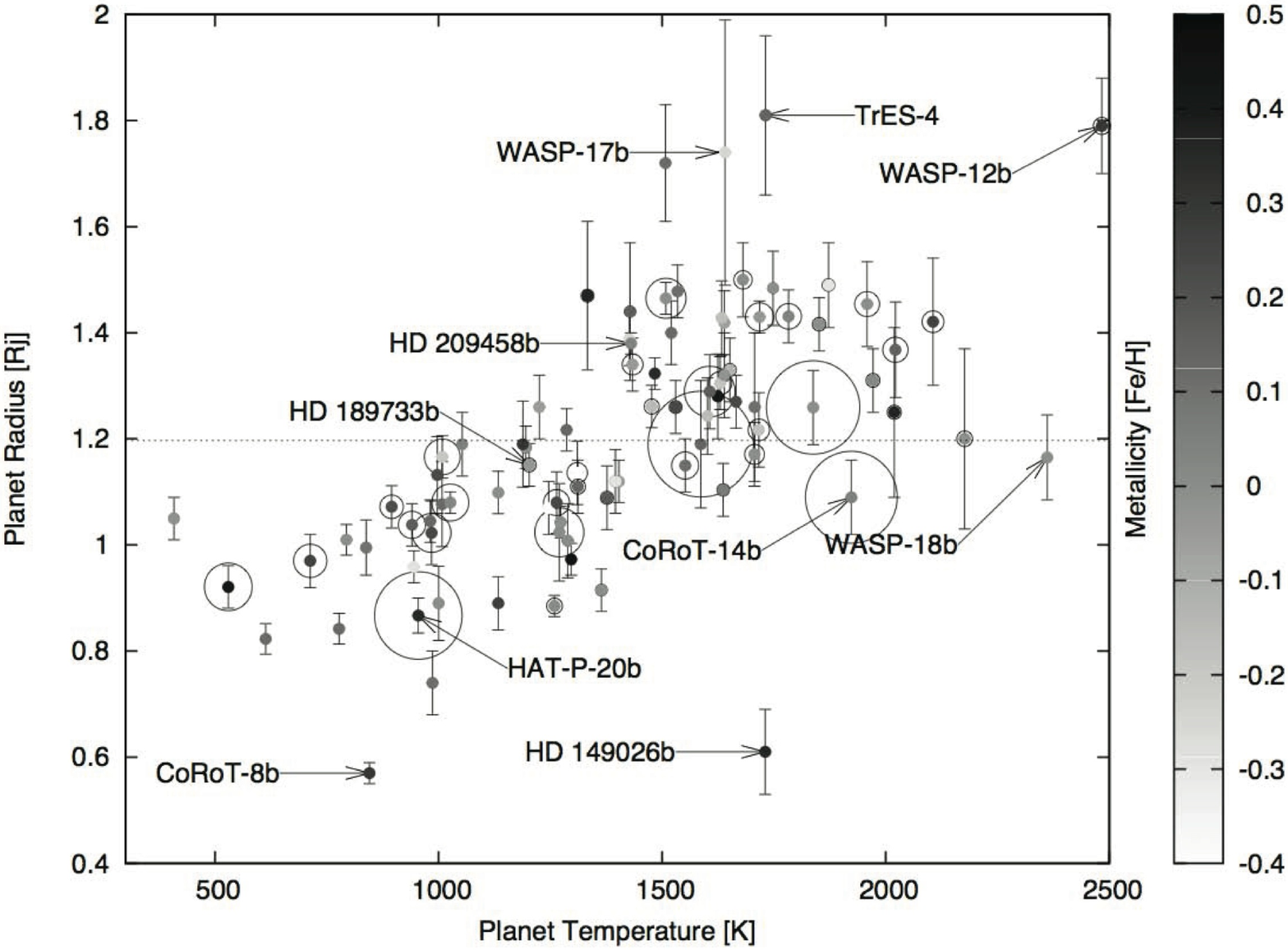}
   }
    \label{Fig1:Temperature vs. Radii}
\caption{Radii and estimated effective temperatures for 90 transiting extrasolar planets with well-determined masses and radii. Circle size is in proportion to planetary mass, and the color of the inner circle indicates metallicity.}
\end{figure} 

\begin{figure}
{
  \includegraphics[scale= .5]{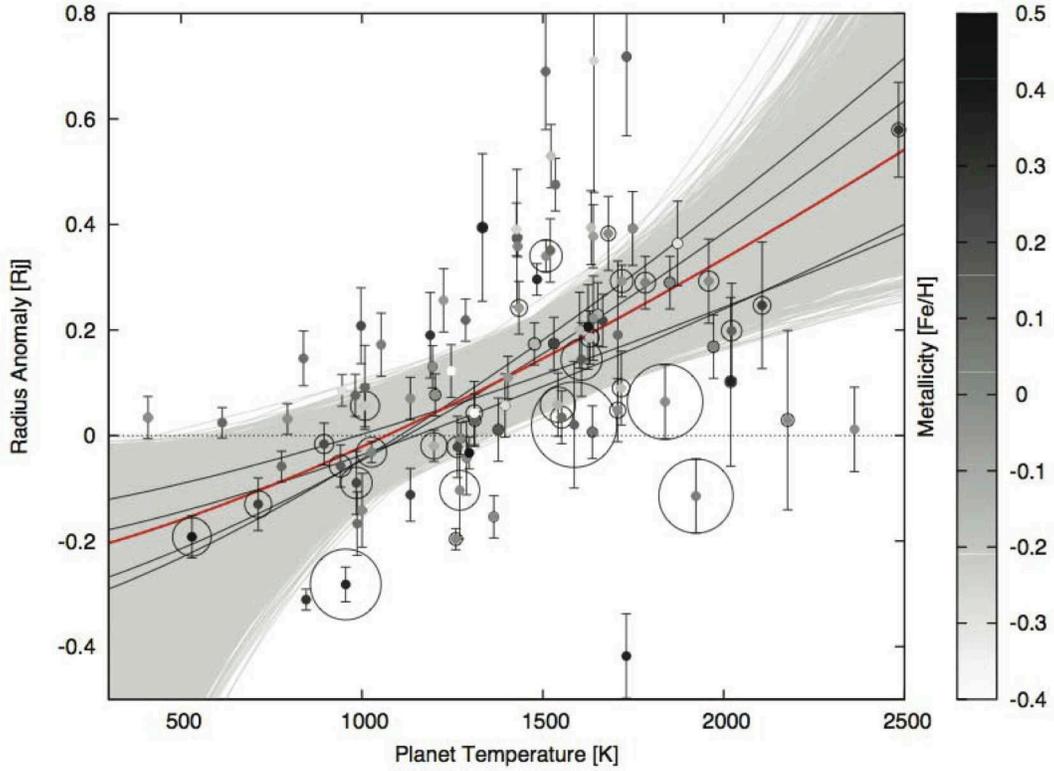}
}
    \label{Fig2:Temperature vs. Radius Anomaly}
\caption{Radius anomaly, $\cal R$, vs. planetary effective temperature, $T_{\rm eff}$, (see Eqn [1]) for 90 transiting extrasolar planets. The red line charts the best error-weighted power-law fit  to the data (${\cal R}\propto T^{\alpha}$, with $\alpha=1.4$). The 10,000 light gray lines show analogous best-fit power-laws to bootstrapped data sets in which the contributing planets are redrawn with replacement from the original data. The $5^{\rm th}$, $15^{\rm th}$, $85^{\rm th}$ and $95^{\rm th}$ percentile bootstrap fits are indicated with black lines. We have adopted the $15^{\rm th}$ and $85^{\rm th}$ percentile fits as an estimate of the 1 $\sigma$ confidence limits on $\alpha$, where we find $\delta \alpha \approx 0.6$. For each planet plotted, the size of the associated variable circle is proportional to planetary mass. The gray-scale color of the inner circle indicates host-star metallicity.}
\end{figure}

\begin{figure}
{
  \includegraphics[scale= .5]{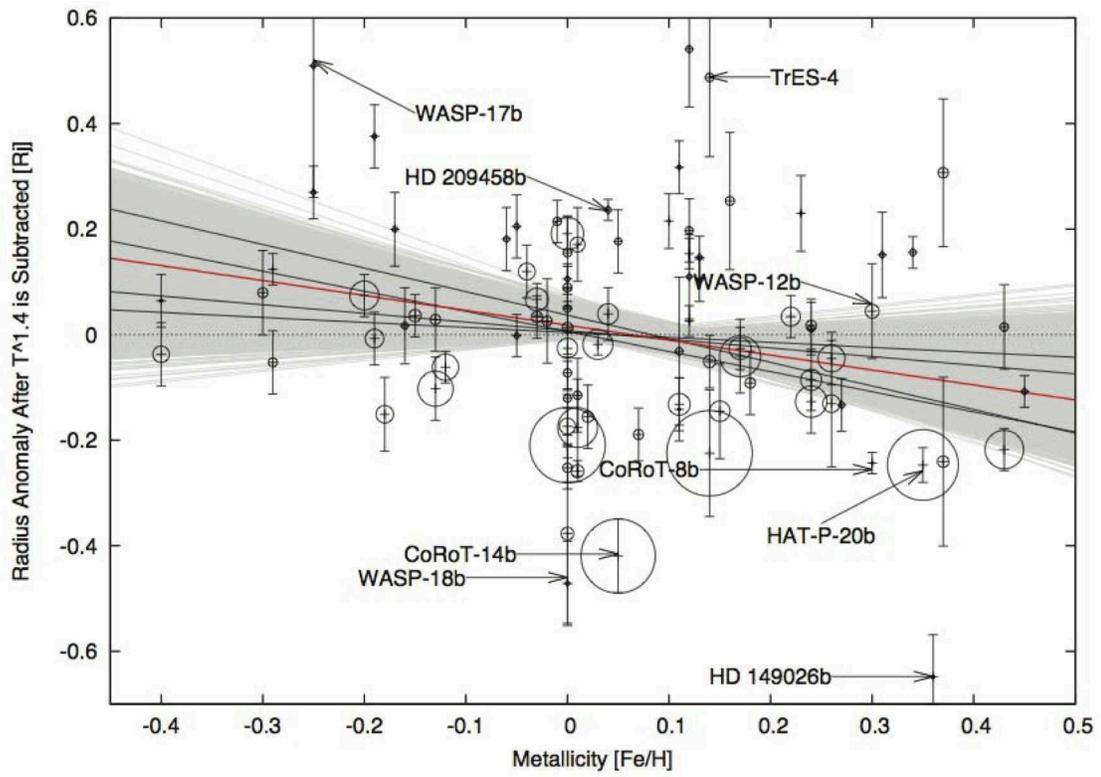}
}
    \label{Fig3:Metallicity vs. Radius}
\caption{Residual radius anomaly vs. host star $\rm{[Fe/H]}$ for 90 transiting extrasolar planets. The best-fit power-law and bootstrap realizations are as in Figure 2.}
\end{figure}

\end{document}